# U.S. National Liquefaction Hazard Maps and their Implications for Engineering Practice and Policy

Morgan D. Sanger[1], Victoria P. Zdanovski[2], and Brett W. Maurer, Ph.D.[3]

## ABSTRACT

This study introduces U.S. national liquefaction hazard maps (NLHMs) developed using a mechanics-informed, geospatial machine learning model which surrogates state-of-practice liquefaction models, exploits a large library of geospatial predictors to infer subsurface conditions, and is anchored to measured conditions with in-situ test data. By convolving this geospatial liquefaction model with the 2023 U.S. national seismic hazard model, liquefaction hazard is mapped across the contiguous U.S. at ~90 m resolution within both conditional (2,475-year design event) and unconditional (return period of ground failure) formulations using high-performance computing for the high-resolution magnitude-disaggregation. The resulting NLHMs provide insights for land-use policy, preliminary site assessment, regional-scale earthquake simulation and response planning, and screening tools for regulatory enforcement, among other applications. Beyond quantifying and visualizing liquefaction hazard, the NLHMs are used herein to examine three questions of engineering practice and policy across a continuous spatial domain: (i) the effect of selecting modal versus mean magnitude in conditional analyses; (ii) the differences between conditional and unconditional hazard formulations; and (iii) the extent to which liquefaction hazard compounds with socioeconomic vulnerability. Results elucidate where and how the choice of magnitude alters computed hazards; that unconditional maps reveal important spatial deviations suppressed by single-scenario maps, which are convenient and widely used in current building codes, but less than completely rational; and that modest but statistically significant socioeconomic gradients in liquefaction exposure exist. The NLHMs should not be used in lieu of site-specific analyses nor to supplant intensive region-specific efforts, but rather provide a baseline prediction of liquefaction hazard, developed by a standard approach, for the contiguous U.S.

[1] Graduate Student, Dept. of Civil and Environmental Engineering, University of Washington, Seattle, WA, USA, sangermd@uw.edu, Corresponding author

[2] Geotechnical Engineer II, HWA GeoSciences Inc., Bothell, WA, USA, VZdanovski@hwageo.com

[3] Associate Professor, Dept. of Civil and Environmental Engineering, University of Washington, Seattle, WA, USA, bwmaurer@uw.edu

## INTRODUCTION

Probabilistic seismic hazard analysis (PSHA) forms the basis of seismic design codes and provides a communication framework among scientists, engineers, insurers, policymakers, and the public. PSHA quantifies seismic risk as the occurrence of ground-motion parameters (e.g., peak ground acceleration, *PGA*) within a given return period (e.g., 2,475 years). Assuming earthquakes are "Poissonian" (i.e., independent), these values may also be expressed as having some annual rate or some exceedance probability in some timespan (e.g., a 2,475-year return period equates to a 2% probability in 50 years). In this way, PSHA translates complex data into actionable insights, and national-scale seismic hazard maps resulting from PSHA are known by experts and laypersons alike. By contrast, analogous national maps communicating the hazard posed by soil liquefaction do not exist. This introduction presents the utility of such maps, new regional-scale liquefaction models that facilitate their development, and the questions of practice and policy these maps can be used to investigate.

Liquefaction hazards are routinely studied in site-specific detail and are sometimes mapped across municipalities or regions for the purposes of land-use planning, disaster simulation, insurance estimating, and regulatory enforcement (e.g., among many, Holzer et al., 2006, 2009; Baise et al., 2006; Cramer et al., 2008, 2017; Hayati and Andrus, 2008; Maurer et al., 2014; Sharifi-Mood et al., 2018). State-wide liquefaction zonation maps exist in Washington, for example, and have been used to plan emergency response (e.g., DHS, 2019) and screen construction permit applications (e.g., SDCI, 2023). Yet these and many other regional-scale maps provide only coarse classifications of hazard (e.g., "Low" versus "High") and do not consider seismic demand. In effect, the maps predict classifications of liquefaction susceptibility, usually inferred from surficial geology maps, but do not explicitly predict liquefaction occurrence or consequences for any earthquake scenario or return period. In California, the 1990 Seismic Hazard Mapping Act required state geologists to delineate zones where liquefaction may occur. Landowners must disclose if their parcel resides in such a zone during real-estate transactions. Yet as of our writing, only portions of the San Francisco Bay and Los Angeles regions have been mapped (CGS, 2024), leaving ~95% of the state unevaluated. Specific to residential structures, one- and two-story homes (i.e., 90% of the U.S. building stock according to Potter, 2022) are typically under the purview of the International Residential Code (e.g., ICC, 2021), which makes no mention of liquefaction. This means any requirement to consider or disclose liquefaction hazard is at the discretion of local officials. This state-of-practice in the U.S., and the paucity of informative hazard maps, has likely led to liquefaction being overlooked for residences in many locations and, in turn, to uninsurance or underinsurance. It is estimated, for example, that less than 20% of homeowners in Oregon and Washington have earthquake insurance (Doughton and Gilbert, 2018). Regardless of the type of structure, the lack of spatially continuous predictions of liquefaction makes it challenging to accurately account for this hazard when forming land-use policy, simulating earthquakes and planning response, prioritizing retrofits within portfolios of distributed infrastructure, or creating screening tools for regulatory enforcement. We attribute the lack of U.S. national liquefaction hazard maps, or NLHMs, to two factors.

*First*, state-of-practice liquefaction models require subsurface data, such as those from the cone penetration test (CPT), but these data cannot feasibly be collected continuously across regional scales. Consequently, regional-scale studies often employ models based on "geospatial" proxies of subsurface conditions, including metrics of topography, geology, hydrology, geomorphology, ecology, and climate. Using these proxy variables, geospatial liquefaction models (GLMs) infer the makeup and response of subsurface soils without subsurface explorations. Although simple GLMs have existed for decades (e.g., Youd and Perkins, 1978; NIBS, 1997), the adoption and acceptance of these models for certain use cases is increasing along with the sophistication of the models, spurred by growth in field observations, geospatial data, and algorithmic learning (e.g., machine learning, ML). Recent studies that train, test, or employ GLMs include Rashidian and Baise (2020), Geyin et al. (2020), Bozzoni et al. (2021), Lin et al. (2021), Todorovic and Silva (2022), Jena et al. (2023), Geyin et al. (2022), Kim (2023), Bullock et al. (2023), Asadi et al. (2024), and Azul et al. (2024). All these efforts have grown the science of liquefaction modeling at regional scales, yet most GLMs suffer from one or more limitations. Namely, they: (i) predict liquefaction without learning from, or anchoring to, state-of-practice triggering and manifestation models (e.g., CPT-based models) that reflect knowledge of liquefaction mechanics; (ii) are not updated by local geotechnical data, even though such data may be prevalent and contradict the inferences of the geospatial model; (iii) use relatively little of the publicly available geospatial information; (iv) are trained by traditional statistical methods (e.g., logistic regression) that cannot capture complex nonlinear relationships or exploit large libraries of geospatial data; and (v) are not provided as code or executable software, meaning it is difficult to impossible for anyone to use the model. Thus, the perception may be that current GLMs are insufficient or unavailable to develop useful NLHMs.

*Second*, the development of an NLHM at high spatial resolution (say, the size of a building parcel, or that of a GLM) requires execution of a PSHA model at hundreds of millions of locations. The contiguous 48 U.S. states, for instance, comprise 889 million cells at ~90 m resolution when considering only relatively flat ground. And, because state-of-practice liquefaction models require both *PGA* and earthquake moment magnitude (*M*) as inputs, magnitude disaggregation of the *PGA* hazard curve is necessary, growing the computational burden. We estimate, for example, that executing the U.S. national seismic hazard model (Petersen et al., 2024) in the manner, extent, and resolution above could require 30 years of serial runtime on a typical PC for just one return period of shaking. Moreover, because liquefaction can be induced by shaking over a wide range of return periods, a *true* depiction of hazard necessitates consideration of the full return-period domain, meaning the PSHA should be executed at numerous return periods for every location.

In this study we develop U.S. NLHMs using the GLM of Sanger et el. (2025), a mechanics-informed ML model that surrogates geotechnical liquefaction models, and which was trained on over 37,000 CPTs from 48 U.S. states and 19 countries. This model has conceptual benefits over prior GLMs in that predictions: (i) are anchored to geotechnical models and the knowledge of liquefaction mechanics embedded therein; (ii) use ML to exploit a large library of predictive information; (iii) are geostatistically updated near in-situ tests, thereby ground-truthing the ML predictions; and (iv) are effectively precomputed everywhere on earth for all possible earthquakes (Sanger et al., 2025). This GLM is coupled herein with high-performance computing (HPC) to execute and disaggregate the 2023 national seismic hazard model at high spatial resolution, over a wide range of return periods, for the entire contiguous U.S., resulting in what we believe to be the first U.S. NLHMs.

While the NLHMs themselves are useful for quantifying and visualizing liquefaction hazard across the U.S., they also provide a means to investigate questions of practice and policy, three of which are posed in this study. The *first* relates to the choice of $M$ used in conventional liquefaction hazard analyses, for which the result is conditioned on the occurrence of a specified design-level earthquake (e.g., the 2,475-year return-period event). Although building codes such as ASCE 7-22 (ASCE, 2022) define a conditional event in terms of shaking intensity (e.g., the 2,475-year $PGA$), they do not explicitly specify whether the associated $M$ be taken as the mean, modal, or some other $M$ from PSHA disaggregation. This ambiguity has led to inconsistent guidance and different selections by practicing engineers, both in the U.S. and internationally (e.g., Franke et al., 2016, 2019; Green and Kizer, 2020; Barani et al., 2023; ODOT, 2024). Although prior studies have shown at point locations the sway this decision can have on liquefaction predictions, we elucidate over a continuous national domain how the choice of either mean or modal magnitude interacts with local conditions to alter the computed liquefaction hazard.

The *second* question concerns the difference between conditional and unconditional hazard formulations. Although the conditional approach is familiar, conveniently simple, and easy to communicate, it is not completely rational, as it represents only one scenario from a continuous distribution of possible events. Weaker motions that occur more frequently, and stronger motions that occur more rarely, can also induce liquefaction. Thus, the unconditional approach integrates over the full spectrum of possible ground motions to yield the annualized rate, or return period, of liquefaction in terms of triggering, surface manifestation, or some other consequence. This approach is very seldom used in practice, but it has clear and compelling advantages (e.g., Kramer and Mayfield, 2007; Ulmer and Franke, 2016; Makdisi and Kramer, 2024; Maurer et al. 2025; Makdisi et al., 2026). Prior studies have demonstrated the merits of the unconditional approach at point locations, typically using simple hypothetical soil profiles. We contribute to this discussion by comparing the two hazard formulations on a continuous spatial basis while accounting for the expected in-situ soil and groundwater conditions.

The *third* question explores how liquefaction hazard patterns relate to population density and socioeconomic exposure. Overlaying the NLHMs with census tract data enables the first nationwide survey of which communities, and how many people, are exposed to liquefaction hazard. Although socioeconomic correlations with other hazards have been studied, such as air pollution (e.g., Tessum et al., 2019), flood risk (e.g. Wing et al., 2022), and heat islands (e.g., Hoffman et al., 2020), we are unaware of any such work in the context of liquefaction hazard. Collectively, our inquiries using the NLHMs provide insights for engineering decision making, policy formation, and social science.

## DATA AND METHODOLOGY

### *Geospatial Liquefaction Model (GLM)*

The Sanger et al. (2025) GLM couples mechanics-based insights from CPT-based liquefaction models with geospatial data and ML. Rather than predict liquefaction triggering in a certain soil at a certain depth, the GLM instead predicts a soil profile's cumulative liquefaction response, or damage potential, at the ground surface, via one of three manifestation indices (*MIs*): the liquefaction potential index (*LPI*) (Iwasaki, 1978); a modified *LPI*, termed $LPI_{ISH}$ (Maurer et al., 2015); and the liquefaction severity number (*LSN*) (van Ballegooy et al., 2014). Because these *MI*s are well known in the literature and included in popular software (e.g., *CLIQ* by GeoLogismiki), their formulae are omitted here but are given and discussed in the *Supplemental Materials*.

To predict each *MI*, the GLM used 37 geospatial predictors that correlate to subsurface traits relevant to liquefaction susceptibility and response, such as soil typology, saturation, relative density, and thickness. Formally, the *MI* at a given location is expressed as a function of ML-predicted parameters (*A*, *B*) and the magnitude-scaled peak ground acceleration ($PGA_M$):

$$MI = \begin{cases} 0, for\ PGA_M < \frac{A}{100B} g \\ A * (\tan^{-1}(B * (PGA_M - \frac{A/100}{B})^2)), for\ PGA_M \geq \frac{A}{100B} g \end{cases} \quad (1)$$

where *MI* is the manifestation index (i.e., *LPI*, $LPI_{ISH}$, or *LSN*), $\tan^{-1}$ is expressed in radians, $g$ is the gravitational constant, *A* relates to the *MI* value attained at large $PGA_M$, and *B* relates to the rate at which the *MI* increases as a function of $PGA_M$. To train the model, ~37,000 CPTs, mapped in Fig. S1 of the *Supplemental Materials*, were sourced from existing compilations (e.g., New Zealand EQC, 2016; USGS, 2019; Rateria et al., 2024; Regione Emilia-Romagna, 2024) and from newly compiled datasets in North America (Sanger et al., 2024a; Rasanen et al., 2024). Each CPT was subjected to a range of seismic loading, *MI* values were computed by a state-of-practice triggering model, and the resulting $MI$-$PGA_M$ response, which is a signature unique to each site, was fit by Eq. (1). ML models were then trained to predict the fitted *A* and *B* values via the geospatial data at each CPT site, and the trained models were executed worldwide in the forward direction at 0.000833˚ (~90 m) resolution, for 13 billion locations, using HPC. Lastly, predictions of *A* and *B* were geostatistically updated near CPTs, an example of which is shown in Fig. S2. The resulting *A* and *B* values for *LPI* are mapped in Fig. 1 for a portion of the Puget Sound region of Washington. The Sanger et al. (2025) outputs also include a variance map that communicates where, and to what degree, the predicted liquefaction response is updated by local geotechnical data, as demonstrated in Fig. S3.

By storing the expected *MI* as two mapped parameters awaiting a $PGA_M$ (e.g., from an event that has just occurred, or a scenario of interest), the model is easily executed via Eq. (1) without HPC or advanced programming (though HPC is still needed in this study due to the national extent and number of return periods at which the PSHA is executed to produce $PGA_M$ values). To extend each *MI* to a consequence prediction, fragility functions conditioned on these *MI*s have been trained on case-history observations to predict certain outcomes, including general surface manifestation (Geyin and Maurer, 2020), pipeline rupture (Toprak et al., 2019), and foundation damage (Maurer et al., 2025). In this paper, we adopt Geyin and Maurer (2020) to compute the probability of "ground failure" (*PGF*), which may be interpreted as the median probability of observing surface manifestations of liquefaction, such as ejecta, settlement, and cracking, at any location within a given map pixel; the function is defined as:

$$PGF|MI = \Phi\left(\frac{ln(MI)-ln(\theta)}{\beta}\right) \tag{2}$$

where Φ denotes the cumulative normal distribution; $\beta$ is the logarithmic standard deviation (e.g., 1.436 for *LPI*); and $\theta$ is the distribution median, or *MI* at which there is a 50% *PGF* (e.g., 6.993 for *LPI*) (Geyin and Maurer, 2020). In summary, the adopted GLM enables high-resolution, regional-scale prediction of liquefaction manifestation, with conceptual and statistical improvements relative to prior GLMs. For complete coverage of its development, performance, limitations, etc., see Sanger et al. (2025).

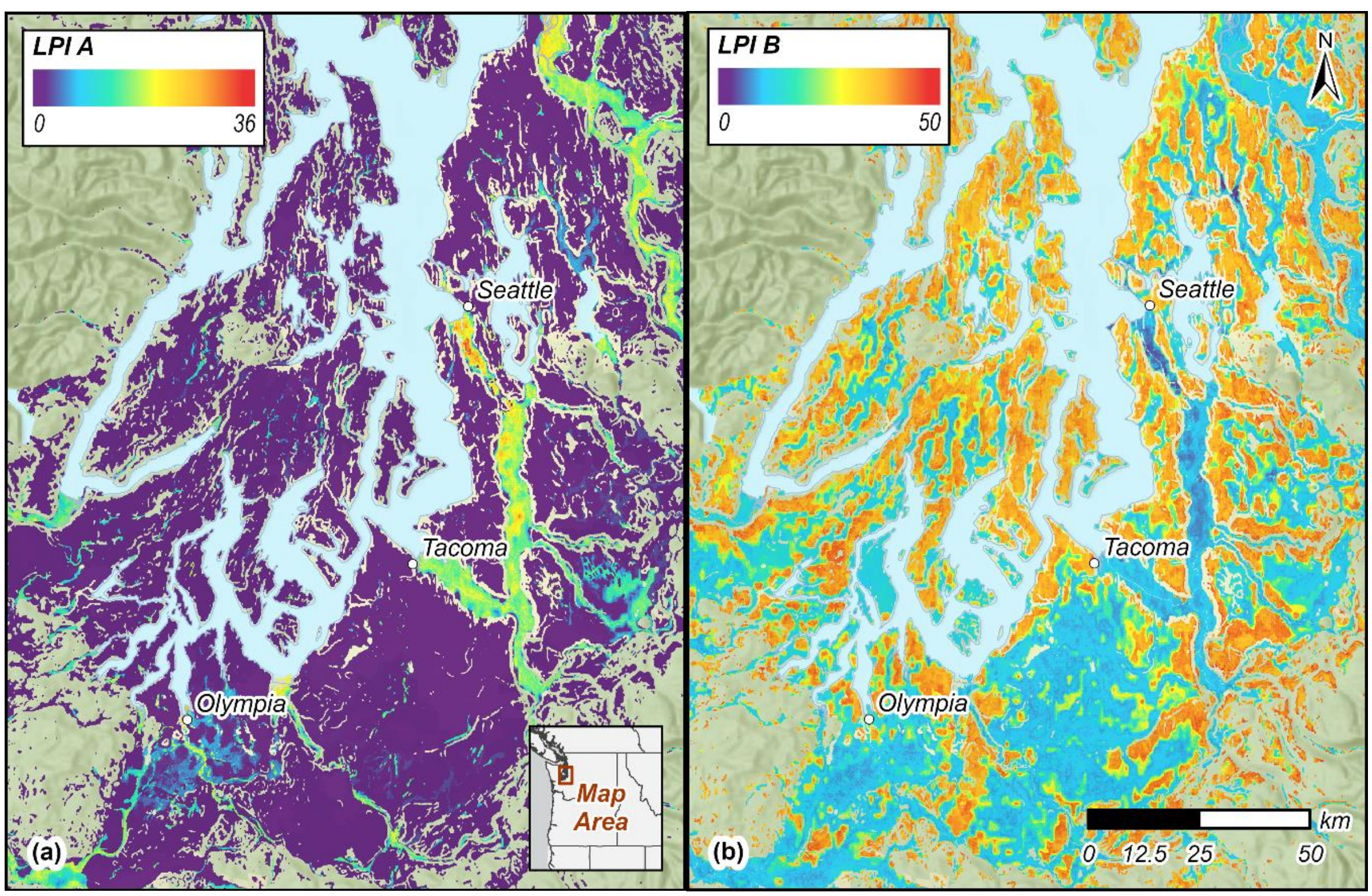

**Fig. 1.** Sanger et al. (2025) GLM products mapped in the Puget Sound region of Washington state: (a) *LPI A* parameter; (b) *LPI B* parameter (Basemap from NOAA NCEI, 2024).

***Conditional and Unconditional NLHM Development***

Using Sanger et al. (2025) to predict *MI* (e.g., *LPI*) and Geyin and Maurer (2020) to predict liquefaction surface manifestations conditioned on *MI*, we map liquefaction hazard within the conditional and unconditional hazard formulations. For the conditional approach, these models are executed using the 2,475-year $PGA_M$, where: *PGA* is obtained from the 2023 national seismic hazard model (Petersen et al., 2024) assuming seismic site class D; *M* is the modal and, alternatively, mean (i.e., weighted average) *M* from disaggregation of the *PGA* hazard curve; and $PGA_M$ is the magnitude-scaled *PGA* computed using the magnitude-scaling factor of Idriss and Boulanger (2008):

$$PGA_M = \frac{PGA}{MSF}, where\ MSF = 6.9\ exp\left(\frac{-M}{4}\right) - 0.058 \leq 1.8 \quad (3)$$

The Idriss and Boulanger (2008) factor is used because the Sanger et al. (2025) GLM was trained to predict *MI* values computed by the Idriss and Boulanger (2008) triggering model; thus, for consistency in forward application, the same factor must be used. Limitations and uncertainties, including those stemming from the adoption of certain models and the assumptions mentioned here, will be discussed in a dedicated section. The conditional modeling workflow is demonstrated in Fig. 2 for the same extent shown in Fig. 1, culminating in a prediction of *PGF* for the 2,475-year earthquake hazard. For brevity, only results using *LPI* are presented and discussed here, but analogous results for $LPI_{ISH}$, *LSN*, and an ensemble of the three models are provided in the *Supplemental Materials* and as a data repository from Zdanovski et al. (2026).

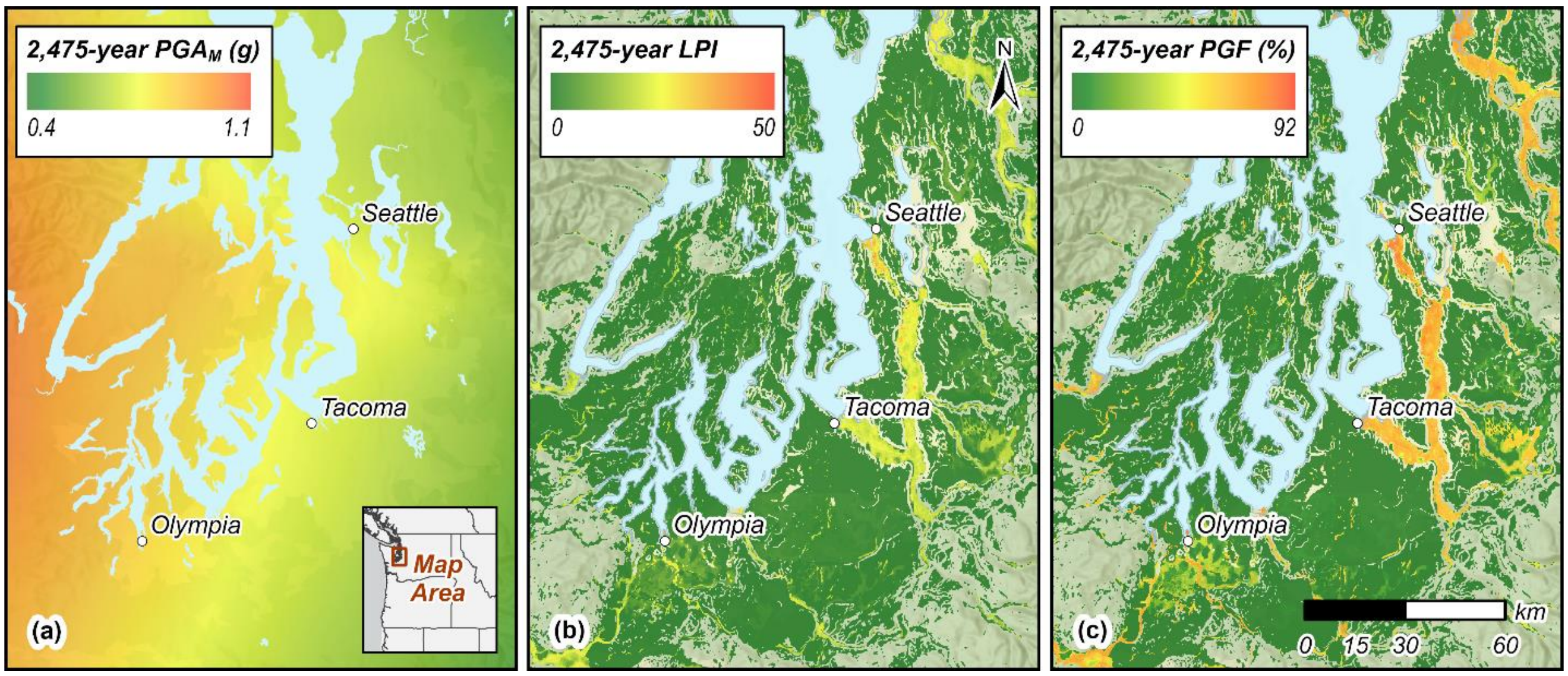

**Fig. 2**. Demonstration of conditional modeling workflow in the Puget Sound region: (a) 2,475-year $PGA_M$ computed using mean *M*; (b) 2,475-year *LPI*; and (c) 2,475-year *PGF* (Basemap from NOAA NCEI, 2024).

In contrast to the approach conditioned on a single return-period event, the unconditional hazard considers every earthquake that could impact a site while also considering the probabilities of those earthquakes occurring. Such an approach is commonly called performance-based earthquake engineering, or PBEE. In its complete form, the PBEE framework computes the mean annual exceedance frequency ($\lambda$) for some decision variable (*DV*) of interest to decision making. This calculation relies on a series of component models and may be generically expressed as:

$$\lambda_{dv} = \sum_{k=1}^{N_{DM}} \sum_{j=1}^{N_{EDP}} \sum_{i=1}^{N_{IM}} P[DV > dv|\, DM = dm_k] \times P\big[DM = dm_k|EDP = edp_j\big] \times P[EDP = edp_j\big|IM = im_i]\Delta\lambda_{im_i} \quad (4)$$

where *DM* is damage measure, *EDP* is engineering demand parameter, *IM* is intensity measure, and where $N_{DM}$, $N_{EDP}$, and $N_{IM}$ are the number of increments of *DM*, *EDP*, and *IM*, respectively. In this context, the *DM* and *EDP* are intermediate variables that link the seismic hazard (i.e., *IM*) to loss (i.e., *DV*). By truncating Eq. (4), the mean annual exceedance frequency of some damage measure, $\lambda_{dm}$, or engineering demand parameter, $\lambda_{edp}$, may alternatively be computed. The latter, for example, is:

$$\lambda_{edp} = \sum_{i=1}^{N_{IM}} P[EDP > edp|IM = im_i]\Delta\lambda_{im_i} \quad (5)$$

Notably, the inverse of $\lambda$ is the return period, which could be computed for any *IM*, *EDP*, *DM*, or *DV* using the PBEE framework. In the context of the NLHMs, *LPI* is effectively a "liquefaction demand parameter" that characterizes the cumulative response, or damage potential, of a liquefiable soil profile. *LPI* therefore relates seismic hazard to the potential for liquefaction-induced damage. $\lambda_{LPI}$ is computed at a given site by convolving the Sanger et al. (2025) GLM with the *PGA* hazard curve, which describes the mean annual rate of *PGA* exceeding a given value ($\lambda_{PGA}$). However, to compute $\lambda_{LPI}$, the *PGA* hazard-curve must be disaggregated by *M*, given that both *PGA* and *M* are required inputs to the GLM (and to any state-of-practice liquefaction model). $\lambda_{LPI}$ is then computed as:

$$\lambda_{LPI} = \sum_{j=1}^{N_M} \sum_{i=1}^{N_{PGA}} P(LPI > lpi|pga = pga_i, m = m_j)\Delta\lambda_{pga_i,m_j} \quad (6)$$

where $N_M$ and $N_{PGA}$ are respectively the number of *M* and *PGA* increments into which the seismic hazard is subdivided, and $\Delta\lambda_{pgai,mj}$ is the incremental annual-exceedance rate for intensity measure, $pga_i$, and magnitude, $m_j$, which is an approach first performed by Kramer and Mayfield (2007); and $P(LPI > lpi|pga = pga_i, m = m_j)$ is the probability that the *LPI* exceeds some value, *lpi*, conditioned on *PGA* and *M*. Calculation of $\lambda_{LPI}$ has previously been shown in the literature (e.g., Goda et al., 2011; Green and Kizer, 2020; Maurer et al., 2025), aside from using the Sanger et al. (2025) GLM to do so. To illustrate, a series of $\lambda_{LPI}$ curves are plotted in Fig. 3 using the 2023 national seismic hazard model, assuming seismic site class D. In Fig. 3(a), $\lambda_{LPI}$ curves are shown for Seattle, Washington for a site where the GLM *A* parameter (see Eq. 1) varies, and the *B* parameter is 15. As liquefaction hazard (i.e., *A*) decreases, $\lambda_{LPI}$ diminishes for any given *LPI*. If *A* is 5, for example, the *LPI* with a 1,000-year return period ($\lambda_{LPI}$ = $10^{-3}$) is 6.5, whereas if *A* is 25, the *LPI* with this return period is 32.5. In Fig. 3(b), $\lambda_{LPI}$ curves are shown for cities in a range of seismic settings, given a site where the GLM *A* and *B* are 25 and 15, respectively, indicative of a high liquefaction hazard. As seismic hazard decreases, so does $\lambda_{LPI}$. In Seattle, San Francisco, California, and Memphis, Tennessee the resulting $\lambda_{LPI}$ curves converge near a 10,000-year return period ($\lambda_{LPI}$=$10^{-4}$), which is to say the 10,000-year $PGA_M$ values are similar in these cities, but for all other return periods the $\lambda_{LPI}$ curves are different. This alludes to the fact that basing decisions on the hazard at one return period, as is ubiquitous in current building codes, can result in locations having the same perceived hazard when the "actual" hazards – considering all return periods – are very different. For every return period less than 10,000 years, $\lambda_{LPI}$ is greater in San Francisco than in Seattle, and greater in Seattle than in Memphis. For this reason, the return periods of ground failure are quite different in the three cities, as discussed and computed below.

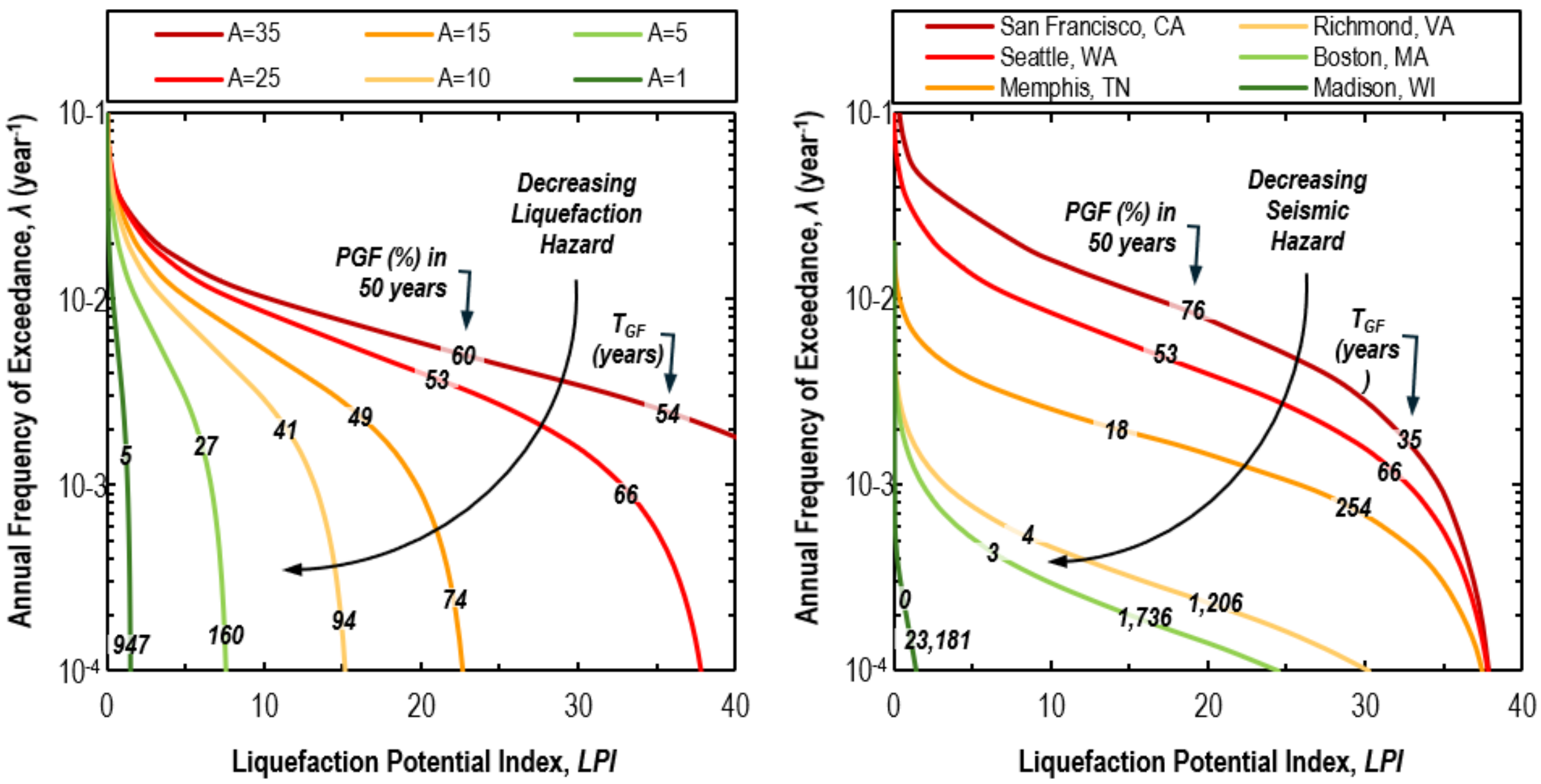

**Fig. 3**. Example *LPI* hazard curves: (a) a site with varying *A* and *B* = 15 in Seattle, Washington; (b) a high liquefaction hazard site (*A* = 25, *B* =15) located in a range of seismic settings.

Having developed $\lambda_{LPI}$ curves at a site, the mean annual exceedance frequency of ground failure, $\lambda_{GF}$, is computed by truncating the PBEE framework prior to the calculation of a *DV*:

$$\lambda_{GF} = \int_{LPI=0}^{\infty} PGF|LPI \cdot \left|\frac{d\lambda_{LPI}}{dLPI}\right| \cdot dLPI \quad (7)$$

where $PGF|LPI$ is defined in Eq. (2); and $\lambda_{LPI}$ is defined in Eq. (6). Although not intuitive, the terms by which $PGF|LPI$ is multiplied in Eq. (7) equate to the probability of a given *LPI* value. Finally, the return period of ground failure, $T_{GF}$, is computed as:

$$T_{GF} = 1/\lambda_{GF} \tag{8}$$

where $\lambda_{GF}$ is defined in Eq. (7). Using this approach, $T_{GF}$ is computed for each $\lambda_{LPI}$ curve in Fig. 3 and added as a label. In Fig. 3(a), $T_{GF}$ varies in Seattle from 54 years to 947 years for the range of *A* and *B* values plotted. In Fig. 3(b), San Francisco, Seattle, and Memphis have respective $T_{GF}$ values of 35, 66, and 254 years (given the same hypothetical *A* and *B* values), despite the *LPI* with a 10,000-year return period being very similar in the three cities. In Madison, Wisconsin, an ice age may occur before the next manifestation of liquefaction. Assuming earthquakes are Poissonian, the probability of ground failure occurring at least once in a timeframe, *t*, may be computed as:

$$PGF(t) = 1 - e^{-t/T_{GF}} \tag{9}$$

where $T_{GF}$ is defined in Eq. (8). Using this approach, the probability of at least one ground failure in 50 years, or $PGF_{50years}$, is computed for each $\lambda_{LPI}$ curve in Fig. 3 and added as a label. Our unconditional NLHM map products include $T_{GF}$, $PGF_{50years}$, and $PGF_{100years}$. Using the same demonstration area as in Fig. 2, these unconditional products are mapped in Fig. 4. Within these map extents, some of the highest hazards are found in Seattle's "South-of-Downtown", or SODO district, where artificial fill and alluvium are prevalent. The computed $T_{GF}$, for example, commonly ranges from 65 to 75 years in much of SODO.

Historically, liquefaction manifestations have been observed throughout SODO during earthquakes in 1949, 1965, and 2001 (Chleborad and Schuster, 1990; Bray et al., 2001; Rasanen et al., 2023). Taking $T_{GF}$ to be 70 years and assuming a 175-year observation period from 1850 (Seattle's founding) to 2025, two occurrences are most likely with 26% probability, followed by three, with 21% probability. Differences between historic observations and predicted return periods would not necessarily discredit the latter, but their relative agreement here nonetheless lends some credibility to the proposed maps. When interpreting Fig. 4, it should be recognized that the return period of liquefaction triggering at-depth in the most susceptible stratum could be different, and likely shorter, than $T_{GF}$, and that the return period of some liquefaction consequences (e.g., foundation damage) could be different, and likely longer, than $T_{GF}$. That is, liquefaction triggering may occur more frequently than some surface manifestation of liquefaction, which in turn may occur more frequently than certain measures of infrastructure damage. Although surface manifestation is a pragmatic and popular proxy of damage potential for infrastructure, it is not a panacea for diagnosing liquefaction effects. Surface manifestation and asset damage due to liquefaction are strongly correlated, but each can occur without the other.

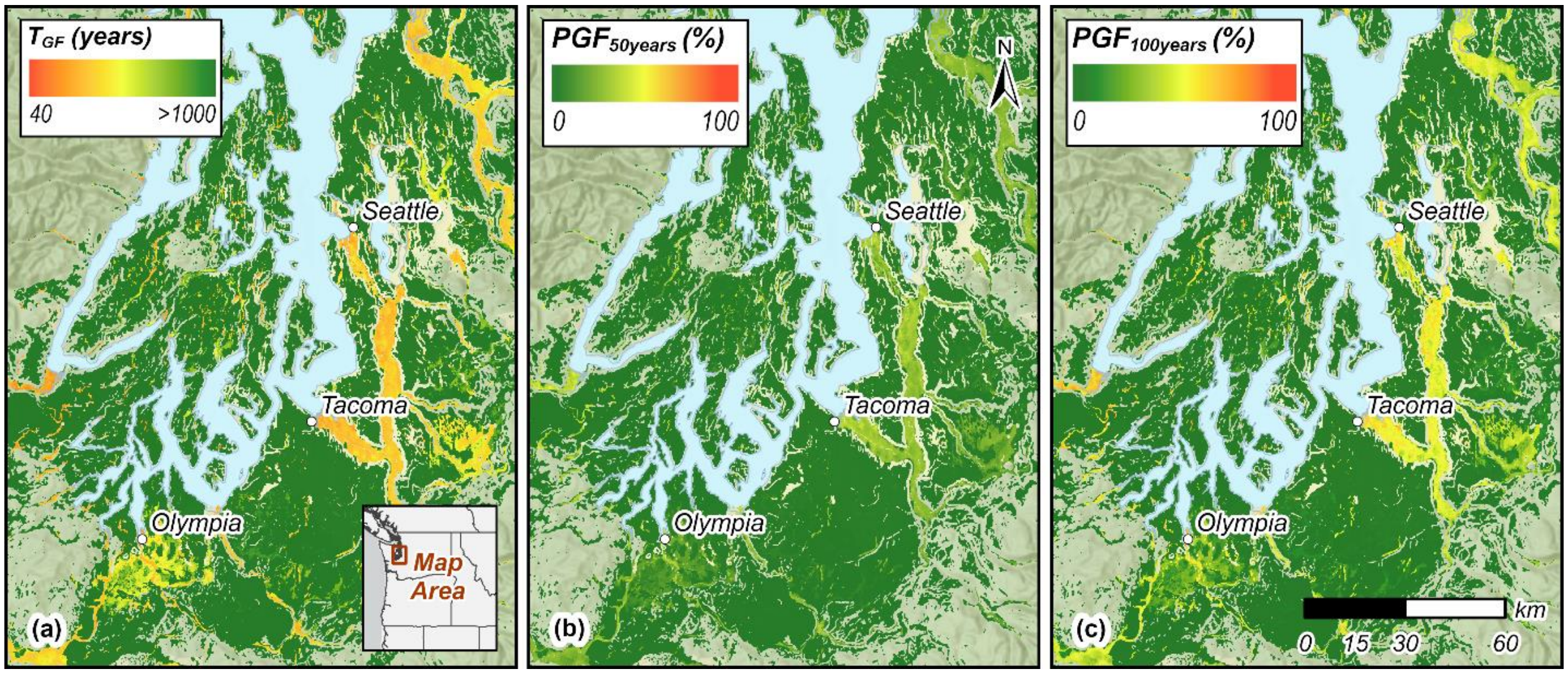

**Fig. 4**. Demonstration of unconditional modeling in the Puget Sound region: (a) $T_{GF}$; probability of one or more ground failures in the next (b) 50 years; and (c) 100 years (Basemap from NOAA NCEI, 2024).

### *Socioeconomic Correlates to Liquefaction Hazard*

To investigate whether computed liquefaction hazards exhibit socioeconomic patterns, we integrate the NLHMs with 150 demographic variables compiled by Esri (2023) at the census-tract level. The roughly 84,000 census tracts have a median area of 4.6 km$^2$ (1.8 miles$^2$) and are generally much smaller in urban areas; 25% of the tracts are less than 1.5 km$^2$ (0.6 miles$^2$), for example, which could equate to the size of a city block. For each tract, indicators of income, housing value, educational attainment, age structure, and employment status are appended to the median 2,475-year *PGF* sampled within each tract, permitting comparison between modeled hazard and the characteristics of affected communities. Linear relationships are quantified using Pearson's correlation coefficient ($r$) and evaluated for statistical significance via two-tailed $p$-values (because the sample is very large, $p$-values are usually ~zero and indicate significance by any benchmark). These analyses are conducted omitting: (i) map cells to which the GLM does not apply – foremost terrain with 90-m resolution slope > 5˚; and (ii) census tracts wherein the median 2,475-year $PGA_M$ is less than 0.1 g, thus constraining the analyses to areas where seismic loading is conceivably sufficient for liquefaction.

Within the remaining ~44,000 tracts, it is also important to further isolate liquefaction hazard from seismic hazard. Because liquefaction is strongly influenced by $PGA_M$, correlations computed without correcting for $PGA_M$ may primarily reflect variations in seismic loading rather than intrinsic differences in site susceptibility or land-use patterns. Analysis of $PGA_M$ alone, for example, shows that high seismic hazard coincides with higher wealth metrics, such as average home value ($r$ = 0.44) and household income ($r$ = 0.15). This is because many of the most expensive coastal and metro corridors, including constrained land markets and those enriched by ports, technology, and finance (e.g., San Francisco, California; San Jose, California; Seattle, Washington, etc.) also have very high seismic hazards. To isolate associations not attributable to seismic hazard, we computed $r(PGF|PGA_M)$, or the partial correlation between the 2,475-year $PGF$ and each variable, while conditioning on the 2,475-year $PGA_M$. Partial correlation removes the effects of $PGA_M$ from both variables prior to evaluating their correlation, enabling direct comparison among locations subject to the same shaking levels. This approach is well-established in statistical analysis (e.g., Kutner et al., 2005) for disentangling the effects of confounding variables.

## RESULTS AND DISCUSSION

All outputs are digitally available from Zdanovski et al. (2026) and shown in Figs. S4-S32. These include: (i) 2,475-year $MI$ and $PGF$ maps, computed using both mean and modal $M$; (ii) a map showing the effect of selecting mean versus modal $M$ on the computed 2,475-year $PGF$; (iii) unconditional map products $T_{GF}$, $PGF_{50years}$, and $PGF_{100years}$; and (iv) the GLM model variance classified by Sanger et al. (2025) which shows where, and to what degree, the mapped values are updated by local geotechnical data and models. In the following, the $LPI$ products are selectively presented and discussed to address the three research questions.

***Mean versus Modal Magnitude in Conditional Hazard Assessment***

Interest in the use of the mean or modal *M* from PSHA disaggregation has long been recognized in general contexts (e.g., Harmsen, 2001; Harmsen et al., 2003) yet in liquefaction assessment, guidance and practices are inconsistent and the issue is controversial. The only guidance appearing in ASCE 7-22 (ASCE, 2022) in the context of liquefaction is that the chosen *M* should be "consistent" with the *PGA*, which is usually interpreted as taking either the mean (i.e., weighted average) or modal *M* from disaggregation. In settings dominated by one seismic source, the two are similar or identical. But in settings with multiple sources, the mean may not represent a realistic scenario; for example, where a hazard is dominated by a nearby small *M* source and distant large *M* source, the mean may correspond to a moderate *M* at moderate distance, but no such source may exist. For this reason, some engineers believe the mean should never be used, and some guidelines explicitly endorse this. The Oregon Department of Transportation geotechnical design manual (ODOT, 2024), for example, states "mean values of *M* and *R* [site-to-source distance] are not recommended for use in liquefaction hazard analyses."

Others have argued that the mean *M* need not represent an actual source to be the choice more consistent with the intents of probabilistic analyses. As said by Kramer (2008): "the fact the mean magnitude itself may not contribute strongly to peak acceleration should not be any more troubling than the fact that 3.5, the expected value of the numbers returned by multiple tosses of a fair die, does not appear on any of the faces of that die." Notably, where the seismic hazard distribution is diffuse, the modal *M* represents a real source, but like any one face of the die, contributes modestly to the overall distribution (e.g., at some locales the modal *M* contributes less than 5% to the total hazard). Although we are unaware of any code or guideline that compels use of mean *M*, many engineers and studies have chosen it (e.g., Kramer and Mayfield, 2007; Cox and Griffiths, 2010; Makdisi et al., 2026), especially when comparing conditional and unconditional formulations. Meneses and Chang (2019) summarized nine building codes in the U.S. and abroad: one required modal *M*, and the remaining ten either stated that both mean and modal *M* are acceptable or – most often – provided even lesser guidance. Some engineers adopt whichever *M* is larger in the name of conservatism (e.g., Barani et al., 2023). Still others have argued that given the nonlinear relationship between *M* and liquefaction hazard, the correct approach is to assess the liquefaction hazard for all *M* values that contribute to the seismic hazard (in terms of factor of safety, *PGF*, or some other metric) and then weigh each output by the relative contribution of each *M* (Kramer, 2008). What this latter recommendation alludes to is that the unconditional formulation, wherein all *M* contributions are explicitly accounted for, would be more rational and eliminate the need to choose a magnitude altogether.

Putting this aside momentarily, the national 2,475-year *PGF* map is shown in Fig. 5, derived from the 2,475-year *LPI* using mean *M*. After generating the same *PGF* product using the modal *M*, we map in Fig. 6 the difference (Δ*PGF*) resulting from the *M* selection, where positive values indicate the mean *M* is greater, and negative values indicate the modal *M* is greater. Across 889 million flatland cells, around 50% see ~0 change in *PGF*, either because the *PGF*s are zero (10% of cells), or because the mean and modal *M* are the same or very similar. Around 20% of cells are positive (mean hazard greater) and largely between 0 and +10% (1% of values are greater than +10%), with a maximum of +22%. Around 30% of cells are negative (modal hazard greater) and largely between 0 and -10% (2% of values are lower than -10%), with a minimum of -34%. On average, there is a -0.34% downshift in *PGF* using the modal magnitude, with a standard deviation of 2.1%.

Thus, for most sites, the choice of *M* translates to *PGF* swings less than ±2.5%, but for select areas it can alter the perceived hazard more meaningfully. The areas most affected tend to have: (i) high liquefaction susceptibility; (ii) *PGA* near thresholds for liquefaction; and (iii) diffuse seismicity – often dominated by areal, or "grid" sources – where multiple earthquakes contribute to the hazard. In these areas, both large positive and large negative Δ*PGF* values may occur in proximity. Whereas the mean *M* weighs the entire seismic hazard distribution and tends to smoothly vary in space (Fig. S4), the mode can be sensitive to minor fluctuations in PSHA inputs and to decisions made in the disaggregation (e.g., magnitude-distance bin sizes; the handling of uncertainty from ground-motion models). For this reason, the modal *M* can spatially vacillate between distinctly different seismic sources, as shown nationally in Fig. S5.

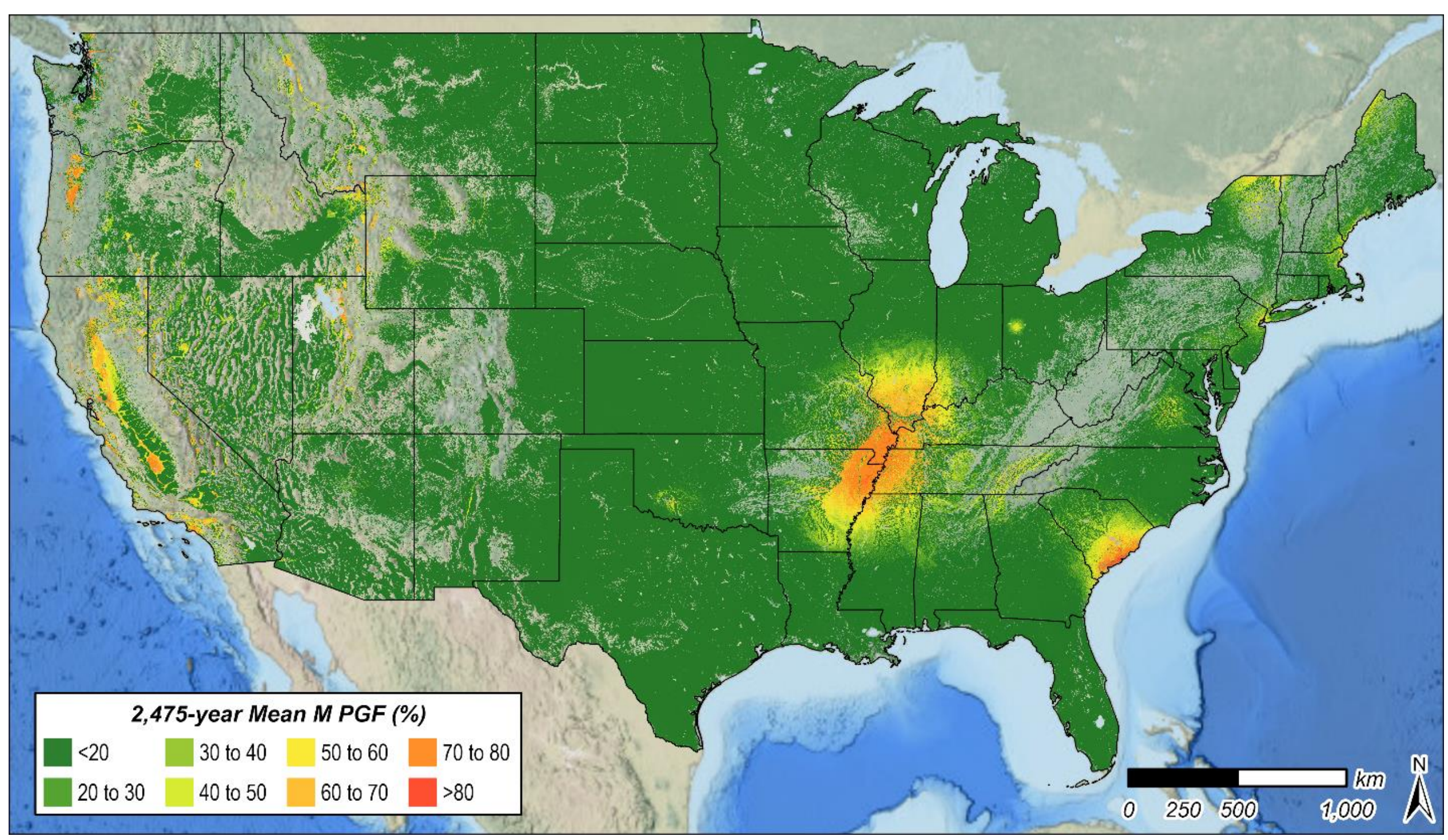

**Fig. 5**. 2,475-year *PGF* (%) computed using mean *M* (Basemap from NOAA NCEI, 2024).

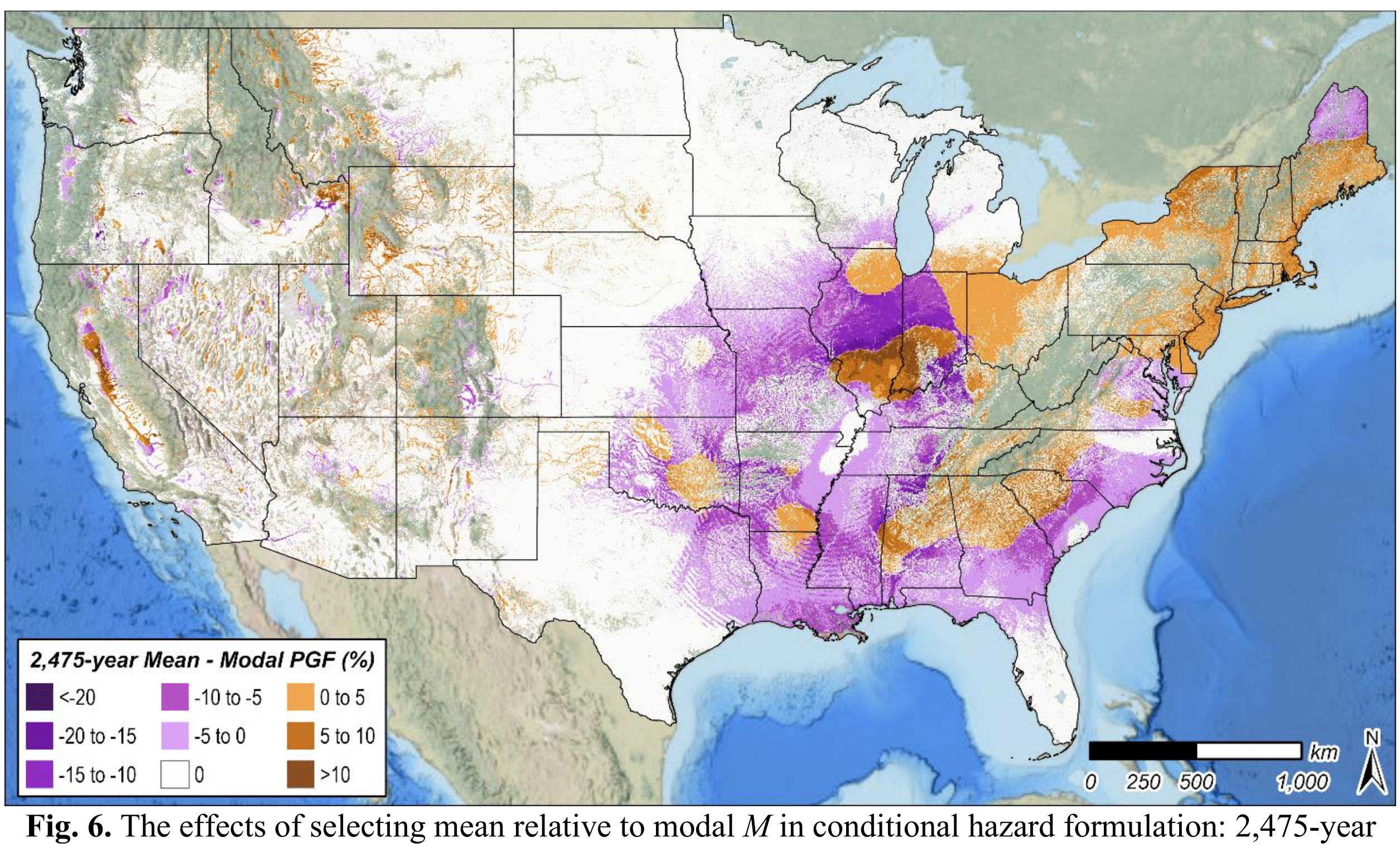

**Fig. 6.** The effects of selecting mean relative to modal *M* in conditional hazard formulation: 2,475-year *PGF* computed using mean *M* minus 2,475-year *PGF* computed using modal *M* (Basemap from NOAA NCEI, 2024).

Near St. Louis, Missouri, for example, the *PGA* hazard distribution is strongly bimodal: 70% is low-*M* grid seismicity and 30% is associated with the high-*M* New Madrid seismic zone. At one location (38.820, -90.653) the mean and modal *M* are 6.4 and 4.9, respectively, where the mode is associated with a local grid source. Nearby (38.879, -90.674), the mean *M* is unchanged, but the modal *M* is 7.8 and associated with the distant New Madrid source. Similar examples are found in all directions radiating from New Madrid. In California's central valley – near the town of Visalia (36.396, -119.119), for example – 80% of the *PGA* hazard is grid seismicity of ~*M* 5.5, with various finite-fault sources contributing 20%, led by a *M* 7.9 San Andreas fault rupture. Here the mean *M* is 5.7, whereas the mode vacillates between *M* 5.3 and *M* 7.9. A final example, yet more extreme, is found along the Cascadia subduction margin for inland Washington, Oregon, and California. Near Klamath Falls, Oregon (41.920, -121.200), for example, the *PGA* hazard is mostly grid seismicity (~60%), with shallow finite-faults contributing ~20% and Cascadia interface ruptures contributing ~10%. Here, the modal *M* swings between 5.3 and 9.3, whereas the mean is a steady *M* 6.7. It is clearly problematic, if not disqualifying, that the mode can dramatically alter a computed liquefaction hazard from one side of a street to the other, all else being equal.

***Conditional versus Unconditional Hazard Formulation***

Shown in Fig. 7 is the computed $T_{GF}$ at national scale. Following the discussion in Fig. 3, it can be shown that locations with identical 2,475-year hazards have very different "actual" (i.e., unconditional) hazards. Sites near Charleston, South Carolina, and Auburn, Washington, for example, have the same 2,475-year $PGA_M$ (0.77 g) and profiles that are modeled as being very similar (i.e., the same *A* and *B* values). Hence these sites have the same 2,475-year *LPI* (36) and *PGF* (87%), reflecting the same very high liquefaction hazard. Yet the seismic settings are distinctly different, giving rise to different $\lambda_{LPI}$ curves, aside from a common crossing point at 2,475-years. Auburn is subject to shallow crustal events, deep intraslab ruptures, and subduction interface earthquakes, collectively having a wide range of return periods, all potentially large magnitude. Conversely, the seismic hazard in Charleston is dominated by one crustal fault that can produce very strong shaking, but which is believed to have a long return period. As a result, the computed return periods of ground failure are very different: 80 years in Auburn and 386 years in Charleston. These $T_{GF}$ values translate into $PGF_{50years}$ values of 46% and 12%, respectively.

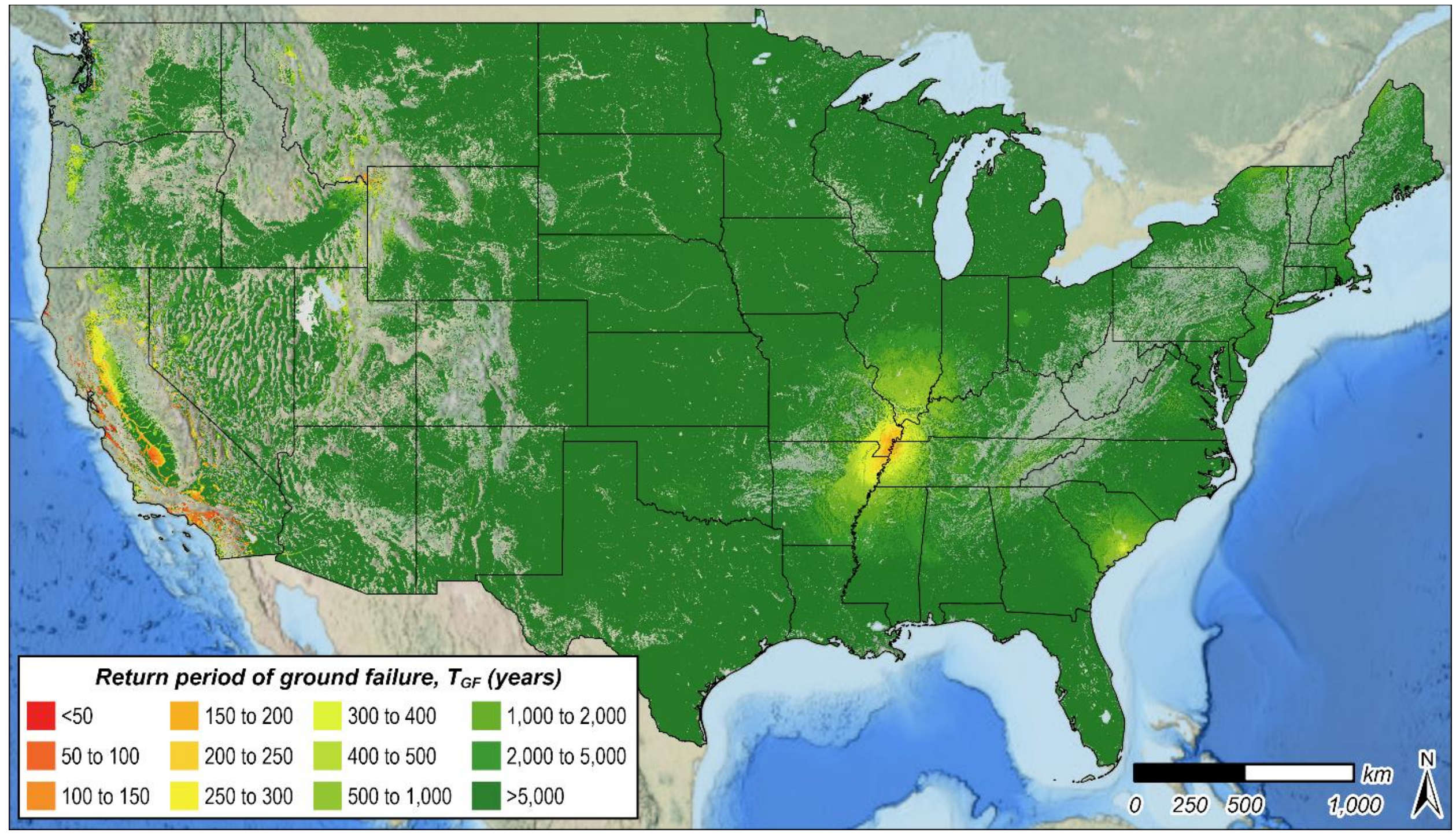

**Fig. 7.** Computed return period of ground failure (years) (Basemap from NOAA NCEI, 2024).

The conditional approach suppresses patterns in the actual liquefaction hazard by ignoring all return periods but one. Yet it is challenging to quantify how the codified hazard might change if it were shifted away from the traditional, conditional formulation, given that the outputs from the two formulations are distinctly different. The conditional approach results in a *PGF* conditioned on the 2,475-year seismic loading (i.e., a single outcome in a single event). The unconditional approach results in a return period of ground failure, which can be expressed as the probability of at least one ground failure over a given exposure time, $t$ (e.g., $PGF_{50years}$ for $t$ = 50 years). Although both approaches output a probability, the comparison is complicated by the different frames of reference: one is a probability given a singular event; the other is a probability over a specific period. If building codes were to adopt unconditional formulations, the acceptable level of risk (e.g., expressed as $T_{GF}$) would need to be defined through careful analysis and community consensus. Just as engineers may presently deem a site "hazardous" (i.e., design unacceptable, mitigation required) when the likelihood of some outcome (e.g., liquefaction triggering, surface manifestation, etc.) in a 2,475-year event exceeds some threshold (e.g., 15%, 50%), an analogous decision threshold is required using the unconditional formulation (e.g., a specified $T_{GF}$, $PGF_{50years}$, or $PGF_{100yeas}$).

A wide range of new decision thresholds could conceivably be justified. One approach demonstrated here is to determine the unconditional threshold that, on average, results in the perceived hazard remaining the same as in the 2,475-year event; specifically, for what time, $t$, does the probability of at least one ground failure, $PGF(t)$, equal the 2,475-year $PGF$? In Fig. 8 we compute and map the exposure time, $t$, that satisfies this condition, making use of the 2,475-year $PGF$ in Fig. 5 and the $T_{GF}$ in Fig. 7. Simply taking the arithmetic mean $t$ across all 889 million flatland cells results in a decision threshold of 400 years (of course, weighting schemes based on population density, liquefaction hazard, etc. can be imagined). Where the mapped, required $t$ is less than 400 years, the unconditional hazard is relatively high compared to the 2,475-year hazard, whereas when the required $t$ is more than 400 years, the unconditional hazard is relatively low compared to the 2,475-year hazard. Again, using the sites from Auburn, Washington and Charleston, South Carolina as an example, the $t$ required in Auburn ($T_{GF}$ = 80 years, 2,475-year $PGF$ = 87%) is 163 years whereas that required in Charleston ($T_{GF}$ = 386 years, 2,475-year $PGF$ = 87%) is 787 years. Adopting the decision threshold of 400 years, engineers would be required to determine the probability of at least one ground failure in 400 years, analogous to how they currently determine and interpret the 2,475-year $PGF$. With this approach, the newly computed hazard, relative to the 2,475-year conditional hazard, would increase in Auburn and decrease in Charleston.

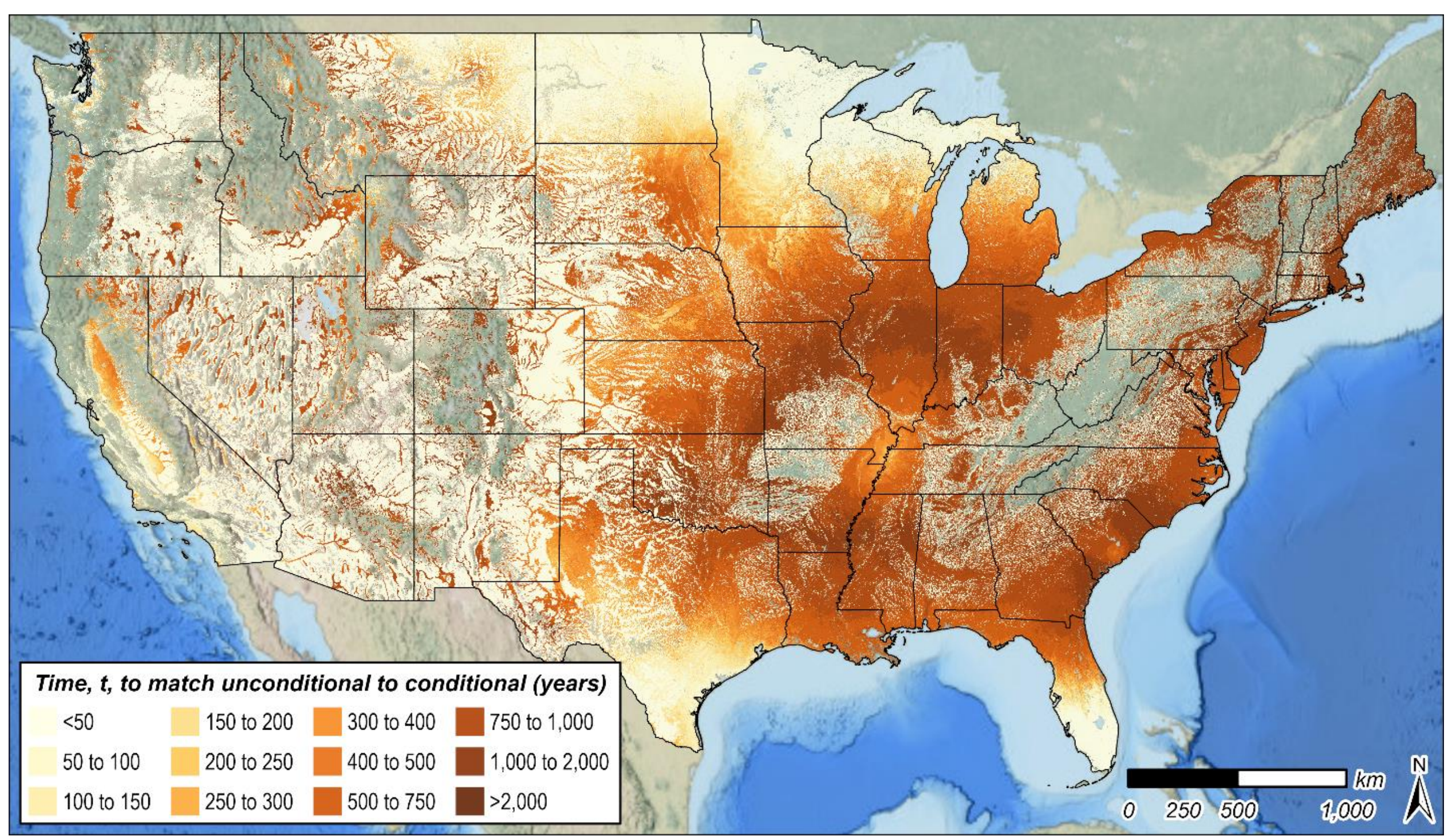

**Fig. 8.** Exposure time, *t* (years) required for unconditional hazard to "match" conditional hazard (Basemap from NOAA NCEI, 2024).

Although limitations will be discussed comprehensively in a dedicated section, we note here that using *LPI* or other *MI*s to quantify liquefaction hazards may not be desirable for all use-cases; an analogous approach centered on liquefaction triggering would undoubtedly produce different results from those mapped in Fig. 8, given that *LPI* transforms $PGA_M$ into a probability of surface manifestation, whereas an alternative framing could transform $PGA_M$ into a probability of triggering in a specific stratum at depth. Each of these approaches could be preferrable in certain contexts. Our example should not be interpreted as an endorsement for framing building codes around manifestation probabilities, but rather as a contribution to the discussion of how unconditional and conditional hazards can differ.

***Socioeconomic Exposure to Liquefaction Hazard***

After controlling for $PGA_M$, residual correlations between liquefaction hazard (2,475-year *PGF*) and several categories of socioeconomic metrics are observed, meaning that liquefaction hazard is not uniformly distributed across communities. Select correlations are given in Table 1, grouped by six categories, and results for all 150 metrics are presented in Table S1 of the *Supplemental Materials*. Across tracts with comparable $PGA_M$, liquefaction hazard has positive correlation with various measures of population density (e.g., $r = +0.19$**)**, meaning that denser, more urban tracts tend to sit on more liquefiable deposits (e.g., river valleys, deltas, fill, waterfronts, industrial corridors.). Tracts with lower homeownership rates and higher mortgage-cost burdens are also associated with higher *PGF*. Before controlling for $PGA_M$, *PGF* is positively associated with high home-value areas due to the underlying positive correlation between seismic hazard and real estate value. Once $PGA_M$ is held constant, however, high *PGF* is more common in lower-wealth, lower-income tracts, while high-wealth enclaves at the same $PGA_M$ have lower *PGF*. In this category, household income, disposable income, the socioeconomic status index, and many high-value home brackets (e.g., $2M+, $1.5M+) have negative *r* (i.e., *PGF* decreases as wealth metrics increase), while very low-income tracts (e.g., household income < $15k) have positive *r* (i.e., *PGF* increases as poverty metrics increase). Consistent with the portrait of wealth and income, tracts with lower educational attainment tend to have higher *PGF*. In this category, the greatest association was measured in tracts where a higher percent of adults have attained less than a 9th-grade education ($r = +0.11$). At the same time, liquefaction hazard is elevated in younger, working-age neighborhoods and reduced in older, senior-dominated areas (e.g., $r = +0.12$); the hazard is also higher where measures of economic polarization are greater (e.g., Gini index, 90%:10% income ratio). Taken together, these results indicate liquefaction exposure is more common in densely populated communities with lower income, lower wealth, and lower rates of educational attainment and homeownership.

**Table 1**. Partial correlations between 2,475-year *PGF* and select variables, controlling for $PGA_M$.

| Category | Variable | $r\ (PGF \mid PGA_M)$ |
|---|---|---|
| Urban Form | Population Density | +0.189 |
| | Daytime Population Density | +0.133 |
| Tenure & Housing Stress | Homeownership Rate | -0.206 |
| | % of Income for Mortgage | +0.103 |
| Wealth, Income, & Socio-Economic Status | Wealth Index | -0.133 |
| | Socioeconomic Status Index (SES) | -0.127 |
| | Median Household Income | -0.102 |
| | Home Value $1.5M-$2M | -0.102 |
| | Disposable Income > $200k | -0.090 |
| | Households with Income < $15,000 | +0.072 |
| Education | % Age 25+ with < 9th Grade Education | +0.112 |
| | % Age 25+ with < High School Degree | +0.111 |
| | % Age 25+ with Bachelor's or Higher | -0.046 |
| Age Structure | Median Age | -0.123 |
| | Senior Population (65+) | -0.093 |
| | Population Age 25-29 | +0.076 |
| Economic Inequality | P90/P10 Income Ratio | +0.093 |
| | Gini Index | +0.087 |

All demographic correlations can be considered moderate or weak per standard benchmarks, but at the population scale can amount to disproportionate exposure for large numbers of people; these correlations are similar in magnitude, for example, to those measured in economic stratifications of air pollution (Collins et al., 2022). These results provide an initial indication that socioeconomic vulnerability may compound with physical hazard exposure in shaping earthquake impacts. Liquefaction has the potential for both short and long-term harm by reducing housing stability, inhibiting recovery, disrupting infrastructure, degrading environmental quality, and exacerbating wealth disparities.

### ***Limitations, Uncertainties, and Future Work***

The NLHMs are subject to limitations and uncertainties that encourage further work. *First*, the sciences of predicting ground motions and liquefaction are imperfect and ever improving. Future versions of the national seismic hazard model, released every few years, could alter the modeled hazard in some areas. Similarly, the in-situ data available to the Sanger et al. (2025) GLM, which are used to train the model and to geostatistically update its predictions in the field, are continually growing. The Sanger et al. (2025) approach is intended to be a "living model" with version updates that reflect new data and the latest science. The triggering and manifestation models surrogated by Sanger et al. (2025) and adopted here will inevitably be supplanted, and new models might alter the mapped liquefaction hazard. The NLHMs are also subject to other GLM limitations discussed in detail by Sanger et al. (2025). For example, the GLM is limited by the spatial resolutions and accuracies of geospatial variables, some of which are themselves predictions. As a result, errors in liquefaction predictions can arise where key inputs, such as modeled groundwater level, are incorrect, where soil conditions vary at a resolution finer than the mapped geospatial variables, or where those variables fail to predict important aspects of the subsurface.

Additional subsurface geotechnical data would inevitably improve reliability. In this regard, CPT soundings are often collected where liquefaction is anticipated or where refusal is less likely. If so, the models could bias toward higher liquefaction probabilities in terrains poorly represented in the training set. Incorporating SPT data in future GLM updates should help strengthen NLHM accuracy in regions or geologic settings where CPTs are scarce (e.g., glacial deposits). Another GLM limitation pertains to lateral spreading. Manifestation models like *LPI* are known not to predict it as efficiently as other expressions of liquefaction (i.e., settlement, ejecta, cracking) (e.g., Maurer et al. 2015), principally because lateral spreading may occur near a free face or on sloping ground, even if the liquefiable stratum is thin. In such a case, the computed *LPI* and *PGF* may be relatively low, but the consequences of liquefaction could exceed those expected in the absence of lateral spreading.

*Second*, all PSHA analyses assumed seismic site class D. If the site class was uniformly different (e.g., E), or if a variable site class were assigned via a $V_{S30}$ model (e.g., Geyin and Maurer, 2023), the NLHM values could change to some degree. However, because *PGA* is used in the GLM (like the state-of-practice models the GLM surrogates), and because *PGA* is less sensitive to site class than lower-frequency spectral accelerations, the changes are often negligible. We performed a sensitivity test in 25 cities having a range of seismic hazards and changed the class from D to E. The change to a softer site generally decreases the *PGA* at longer return periods (for which the rock motion tends to be strong) and increases the *PGA* at shorter return periods (for which the rock motion tends to be weak). In Seattle, for example, the 2,475-year *PGA* changes by -0.048 g while the 50-year *PGA* changes by +0.019 g. Specific to the 2,475-year NLHMs, the hazard mapped throughout the county would generally diminish, albeit typically by a trivial amount. The 2,475-year D-to-E *PGA* changes in select cities include: -0.003 g in Boston, Massachusetts; -0.005 g in Richmond, Virginia; -0.028 g in Memphis, Tennessee; -0.040 g in Portland, Oregon; -0.139 g in San Francisco, California; -0.159 g in Salt Lake City, Utah; and -0.178 g in Los Angeles, California. Notably, the largest changes are associated with large *PGA*s which, when considering Eqs. 1 and 3, cause only a small change in *PGF*. In San Francisco's Marina District, for example, where the mapped liquefaction hazard is among the highest in the city, the -0.159 g shift in *PGA* changes *PGF* by just -0.5%. In Los Angeles, Marina Del Ray has among the highest mapped hazard in the city; there, the -0.178 g change in *PGA* changes *PGF* by -0.8%. In our limited tests, changes to the 2,475-year *PGF* never exceed 1% regardless of the seismic and liquefaction hazards. The effect of the D-to-E change on the unconditional hazard is less straightforward, since *PGA* decreases for some return periods, but increases for others, and since different return periods contribute differently to $T_{GF}$. For the tests in 25 cities, the D-to-E change had a median effect on $T_{GF}$ of +1.64 years, or +0.25%, meaning the hazard diminished slightly. Yet in select locations, larger changes to $T_{GF}$, both up and down, were observed. For example, at a site in Newport, Oregon, with low liquefaction hazard, $T_{GF}$ shortened by 85 years, or -9%, and at a site in Charleston, South Carolina, with high liquefaction hazard, $T_{GF}$ lengthened by 8 years, or +2%. In summary, the NLHMs might, on average, remain the same if site class was changed, but could communicate a meaningfully different hazard in specific places.

*Third*, some compromises between computational expense and accuracy were made within the unconditional approach. PSHA calculations and disaggregation were performed at 0.04˚ (~4 km) spatial resolution, then interpolated to the finer 0.0008˚ (~90 m) resolution of the GLM. Sensitivity tests established that the resulting error was acceptably small and unbiased. Throughout King County, Washington, for example, the ensuing percent error in $T_{GF}$ ranged from -2.53% to +2.38%, with a median of ~0%, thereby justifying the less expensive computation. Additionally, because this approach integrates over *PGA* hazard curves – each with a given return period – it can be sensitive to the number of datapoints used to define those curves (i.e., $PGA$-$\lambda$ pairs), with coarser discretization tending to misrepresent $T_{GF}$. Because PSHA execution at just one return period across the contiguous U.S. requires extensive resources, we assessed the sensitivity of $T_{GF}$ to the quantity of data defining the *PGA* hazard curves in various cities, and for a range of soil conditions (i.e., GLM *A* and *B* values). From these analyses, 20 return periods were found to produce sufficiently accurate results and were used to generate the NLHMs in all locations. Although we find these compromises acceptable for the intended regional-scale purposes of the NLHMs, and necessary considering the computational burden, they do produce small errors.

*Fourth,* the socioeconomic correlation analysis was performed at the census-tract scale using the median modeled liquefaction hazard within each tract and should not be interpreted as applying to individual households. This analysis controlled only for $PGA_M$ and only in a linear sense; it did not account for non-linear relationships or unobserved confounders. The socioeconomic indices are derived from survey data and inevitably contain measurement errors. With 44,000 tracts, very small effect sizes are significant, so emphasis should be placed on the magnitude of relationships rather than *p*-values. Finally, the analysis examined present-day conditions against a long-return-period hazard metric and did not capture how future demographic shifts, land-use change, or adaptation measures may alter the distribution of liquefaction risk. These limitations suggest the correlations should be viewed as hypothesis-generating and complementary to, rather than a substitute for, detailed local studies.

*Finally*, it is emphasized that the NLHMs are not intended to supersede intensive city- or region-specific hazard mapping efforts that may benefit from expert knowledge of local geology, access to larger subsurface datasets, laboratory testing, or more accurate estimates of groundwater depth. Rather, the intention herein is to provide baseline NLHMs, developed by a standard approach, for the entire contiguous US; a future iteration of the NLHMs may include Alaska, Hawaii, and other U.S. territories. Notwithstanding these limitations, this study may be the first effort to develop U.S. NLHMs. As the availability of community geotechnical data increases, so too will the feasibility and accuracy of such attempts. Ultimately, new subsurface data and future earthquake observations will verify or refine the results shared here and succinctly summarized below.

## CONCLUSIONS

By coupling a geospatial liquefaction model with probabilistic seismic hazard analyses, we produced spatially continuous national liquefaction hazard models across the contiguous U.S. using both conditional and unconditional hazard formulations. Potential applications for these models include preliminary site assessment, land-use policy, regional-scale response and recovery planning, prioritization of distributed assets for site-specific study, and screening tools for regulatory enforcement. Beyond these general uses, we investigated questions of engineering practice and policy. The ambiguous choice between mean and modal magnitude in conditional analyses can materially influence predictions in certain settings, altering design-level interpretations of the hazard. Our findings delineate where, to what degree, these alterations are expected and motivate clearer guidance on magnitude selection in design standards. Comparison between unconditional and conditional hazards demonstrates how the latter can obscure the true hazard and result in inconsistent levels of codified risk across the country. Our findings show where these inconsistencies may occur and encourage migration away from conditional hazard formulations in current building codes, even if only incremental, such as that recently proposed by Makdisi et al. (2026). The models enable national-scale evaluation of exposure to liquefaction hazard and socioeconomic metrics, providing preliminary evidence that liquefaction hazard may compound with social vulnerability, shaping post-event mobility, access, and recovery. While not a substitute for site-specific investigation, the products provide a consistent national reference in support of regional-scale applications and inquiries. As geotechnical datasets expand and liquefaction models improve, the presented national liquefaction hazard models can be refined and updated, advancing toward increasingly accurate national representations of liquefaction hazard.

## DATA AVAILABILITY

The geotechnical and geospatial data used in model development are all publicly available, as described and referenced in the text. The Sanger et al. (2025) GLM code and map products are available from Sanger et al. (2024b,c). All map products produced in this study are digitally available on DesignSafe as geotiff files from Zdanovski et al. (2026).

## ACKNOWLEDGEMENTS

The presented work is based on research supported by the U.S. Geological Survey (USGS) under award G23AP00017, the Cascadia Region Earthquake Science Center (CRESCENT) via National Science Foundation (NSF) award 2225286, the Cascadia Coastlines and Peoples (CoPes) Hub via NSF award 2103713, the Pacific Earthquake Engineering Research (PEER) Center under award 1185-NCTRMB, and the Pacific Northwest Transportation Consortium (PacTrans) under award 69A3552348310. However, any opinions, findings, conclusions, or recommendations expressed herein are those of the authors and may not reflect the views of USGS, NSF, CoPes, CRESCENT, PEER, or PacTrans. Furthermore, this work was facilitated using HPC infrastructure provided by the Hyak supercomputer and funded by the University of Washington’s student technology fund, and through DesignSafe at the Texas Advanced Computing Center.

## SUPPLEMENTAL MATERIALS

Figs. S1–S32 and Table S1 are available as a data repository from Zdanovski et al. (2026).

## CONFLICTS OF INTEREST

The authors declare no conflicts of interest.